\input ./style/arxiv-vmsta.cfg
\documentclass[numbers,compress,v1.0.1]{vmsta}

\volume{4}% Updated by VTEXPTS2LaTeX.exe, 30.06.2017 13:05
\issue{2}% Updated by VTEXPTS2LaTeX.exe, 30.06.2017 13:05
\pubyear{2017}
\firstpage{127}% Updated by VTEXPTS2LaTeX.exe, 30.06.2017 13:05
\lastpage{139}% Updated by VTEXPTS2LaTeX.exe, 30.06.2017 13:05
\doi{10.15559/17-VMSTA78}% Updated by VTEXPTS2LaTeX.exe, 03.05.2017
\startlocaldefs
\newcommand{\lleft}{\left}
\newcommand{\rright}{\right}
\allowdisplaybreaks

\newtheorem{prop}{Proposition}

\theoremstyle{definition}
\newtheorem{defin}{Definition}

\endlocaldefs

\begin{document}
\begin{frontmatter}

\title{Multi-state models for evaluating conversion options in~life insurance\tnoteref{t1}}
\tnotetext[t1]{This work is dedicated to Prof. Dmitrii Silvestrov in recognition of his contribution to actuarial
mathematics.}
\author[a]{\inits{G.}\fnm{Guglielmo}\snm{D'Amico}\corref{cor1}}\email{g.damico@unich.it}%\orcid{0000-0002-6948-2912}
\cortext[cor1]{Corresponding author.}
\author[b]{\inits{M.}\fnm{Montserrat}\snm{Guillen}}\email{mguillen@ub.edu}
\author[c]{\inits{R.}\fnm{Raimondo}\snm{Manca}}\email{raimondo.manca@uniroma1.it}
\author[d]{\inits{F.}\fnm{Filippo}\snm{Petroni}}\email{fpetroni@unica.it}

\address[a]{Department of Pharmacy, University ``G. d'Annunzio'' of Chieti-Pescara, Chieti,~Italy}
\address[b]{Department of Econometrics, Statistics and Economics, University of Barcelona, Barcelona, Spain}
\address[c]{MEMOTEF Department, University ``La Sapienza'', Rome, Italy}
\address[d]{Department of Business, University of Cagliari, Cagliari, Italy}

\markboth{G. D'Amico et al.}{Conversion options in life insurance}

\begin{abstract}
In this paper we propose a multi-state model for the evaluation of the
conversion option contract. The multi-state model is based on
age-indexed semi-Markov chains that are able to reproduce many
important aspects that influence the valuation of the option such as
the duration problem, the time non-homogeneity and the ageing effect.
The value of the conversion option is evaluated after the formal
description of this contract.
\end{abstract}


\received{29 March 2017}% Updated by VTEXPTS2LaTeX.exe, 03.05.2017
%10:47
\revised{20 April 2017}% Updated by VTEXPTS2LaTeX.exe, 03.05.2017 10:47
\accepted{21 April 2017}% Updated by VTEXPTS2LaTeX.exe, 03.05.2017
%10:47
\publishedonline{10 May 2017}
\end{frontmatter}

%s1 ###
\section{Introduction}\label{sec1}

The conversion option is an option that allows the policyholder to
convert his original temporary insurance policy (TIP) to permanent
insurance policy (PIP) before the initial policy is due.

Insurance companies may find convenient this kind of contract
because it may be much less expensive to convert the initial policy
instead of issuing a new one. On the other side the policyholder may be
interested in converting the contract because, at the time of
conversion, insurance companies do not require any evidence of
insurability and calculate the new premium according to the age at the
issue of the original contract. However, at the time of conversion the
insured individual has to pay the difference of cash value between the
original TIP and converted PIP.

The literature on conversion option is not large and the main
reference is represented by the article \citep{su10} where a valuation
model was constructed based on mortality tables and then extended to a
Lee--Carter model of mortality. A related article is %that of
\citep{no08} where the author considered an exchange option that is
available in Norway.

In general, insurance companies collect data in form of
sequences of events concerning the health status of the policyholders.
Therefore they can evaluate survival probabilities taking into account
for the health evolution of the insured person. This means that the
adoption of a multi-state model can improve the evaluation process of
policy-linked contracts like the conversion option when compared with
information extracted from simple mortality tables. Indeed, as argued
in \citep{kwjo08}, mortality rates are limited to accurately predict
the dynamics of mortality. Moreover recent literature includes
contributions where multi-state models, based on Markov chains, have
been advanced as a valuable alternative to traditional mortality models
see, e.g., \citep{lili07,lili12,toma91,kwjo06}.

A general approach based on semi-Markov processes has been
applied to problems of disability insurance also in recent years, see
\citep{stmasi07,daguma09,daguma13,ma13}. Their appropriateness is due
to the rejection of the geometric (exponential in continuous time
model) distribution hypothesis for modeling the waiting times in %an
a health status before making a transition in another state. Indeed,
%%from
the geometric (exponential) hypothesis results in
the lack of memory property that is very convenient from a mathematical
point of view but is rarely supported by empirical evidence.

In this paper we focus on the evaluation of the conversion
options when an age-indexed semi-Markov multi-state model describes the
evolution of the health status of the policyholder. To this end we
first derive transition probabilities for the model and then we develop
the evaluation procedure by analyzing the TIP and PIP contracts and the
conversion option. The obtained results %represents
represent the generalization of the results %by
of \citep{su10} in a more general framework. Particularly, we show
that the value of the conversion option depends on many parameters that
are contemporary managed by our model such as the health status
evolution of the policyholder, the age of the policyholder and the
chronological time effect due to medical-scientific progress.

We start in Section~\ref{sec2} by describing the age indexed semi-Markov
model. In \xch{Section~\ref{sec3},}{Section~\ref{sec3}} we explain the valuation procedure of the
conversion option and we calculate its value.
The paper ends %closes
with some conclusions and suggestions for further research.

%s2 ###
\section{Age-indexed semi-Markov model}\label{sec2}

Following the approach of \citep{jama97} it is possible to give a
tractable extension of discrete time non-homogeneous semi-Markov chains
useful to consider different aspects that are relevant for the
evaluation of the conversion option like the duration problem, the
non-homogeneity and the ageing effect. This approach has been further
generalized %by
in \citep{da11,dape11,dape12} where general indexed semi-Markov
processes were investigated and applied to different problems.

On a complete probability space $ (\varOmega, \mathcal{F},
\mathbb{P})$ we consider two sequences of random variables that evolve
jointly:\\
\[
J_{n}:\varOmega\rightarrow E=\{1,2,\ldots,D\},
\]
\[
T_{n}:\varOmega\rightarrow\mathbb{N}.
\]
$J_{n}$ represents the state at the $n$-th transition which can be
identified with one of the mutually exclusive elements of the set $E$.
In our framework, the set $E$ contains all possible values of the
health-status of the policyholder, included the death state denoted by
$D$. The quantity $T_{n}$ denotes the time of the $n$-th transition, i.e.
the time when the policyholder enters in the health-status $J_{n}$.

We define the age-index process by the relation:
%
%e1 ###
\begin{equation}
\label{age-index} A_{n}=A_{n-1}+T_{n}-T_{n-1},\quad n\in\mathbb{N},
\end{equation}
where $A_{0}$ is known. From now on we will set $A_{0}=a$ and as %usual
usually $T_{0}=0$. This implies that by recursive substitution
$A_{n}=a+T_{n}$, that is the age at the time of the $n$-th transition is
given by the initial age ($A_{0}=a$) plus the time of occurrence of the
$n$-th transition ($T_{n}$).

The key assumption is to consider the triple $(J_{n}, T_{n},
A_{n})$ like a non-homoge\-neous Markov Renewal Process with index:
%
%e2 ###
\begin{align}
&\mathbb{P}\bigl[J_{n+1}=j, T_{n+1}\leq t \bigm\vert\sigma(J_{h},T_{h},A_{h}, \, h\leq t), J_{n}=i, T_{n}=s, A_{n}=a+s\bigr]\notag\\
& \quad =\mathbb{P}[J_{n+1}=j, T_{n+1}\leq t \mid J_{n}=i, T_{n}=s, A_{n}=a+s]=\, \,^{a}Q_{ij}(s;t),\label{due}
\end{align}
where $\sigma(J_{h},T_{h},A_{h},\,h\leq t)$ is the natural
filtration of the three-variate process $(J_{h},\break T_{h},A_{h})_{h\in
\mathbb{N}}$.

Relation (\ref{due}) affirms that the knowledge of the values $J_{n},
T_{n}, A_{n}$
is sufficient to give the conditional distribution of the couple
$J_{n+1}, T_{n+1}$ whatever the values of the past variables might be.
Let us denote by $^{a}p_{\mathit{ij}}(s)$ transition probabilities of the
embedded non-homogeneous age indexed Markov chain:
\[
^{a}p_{ij}(s):=\mathbb{P}[J_{n+1}=j\mid J_{n}=i, T_{n}=s, A_{n}=a+s]=\lim
_{t\rightarrow\infty}\,^{a}Q_{ij}(s;t).
\]
Furthermore, it is necessary to introduce the probability that the
process will remain in the state $i$ up to the time $t$ given the
entrance in $i$ at time $s$:
\[
^{a}\overline{H}_{i}(s;t)=\mathbb{P}[T_{n+1}>t \mid J_{n}=i, T_{n}=s, A_{n}=a+s]=1-\sum
_{j\in E}\,^{a}Q_{ij}(s;t).
\]
Now it is possible to define the distribution function of the waiting
time in each state $i$, given that the state successively occupied is known
\begingroup
\abovedisplayskip=5pt
\belowdisplayskip=5pt
\begin{equation*}
\begin{aligned} ^{a}G_{ij}(s;t):=&\,
\mathbb{P}[T_{n+1}\leq t \mid J_{n+1}=j, J_{n}=i,
T_{n}=s, A_{n}=a+s]
\\[-1pt]
 =&\,\lleft\{ %
\begin{matrix} \frac{^{a}Q_{ij}(s;t)}{^{a}p_{ij}(s)} & \textrm{if} \
^{a}p_{ij}(s) \neq0 ,\\
1 & \textrm{if} \  ^{a}p_{ij}(s) = 0 .
\end{matrix} %
\rright. \end{aligned}
\end{equation*}
The main advantage of semi-Markov models as compared to Markovian
models is that in a semi-Markovian environment the probability
distribution functions $ ^{a}G_{\mathit{ij}}(s;\cdot)$ can be of any type. On
the contrary,
in a Markovian model they should be %a
geometrically distributed. Since disability data have shown rejection
of the geometricity of the waiting time distributions (see, e.g. \citep
{hapi99,stmasi07,daguma09,dadijama11}), semi-Markovian models are more
appropriate to describe the
dynamics of \xch{health-status}{health-states} evolution in time.

Let us denote by $^{a}N(t)=\sup\{n\in\mathbb{N}: T_{n}\leq t \mid
A_{0}=a\}$ the process counting the number of transitions up to time
$t$ and define consequently
the age-indexed semi-Markov chain by
\[
^{a}Z(t)=J_{\,^{a}N(t)}.
\]
In the valuation procedure it will be useful to introduce the backward
recurrence time process $B(t)=t-T_{\,^{a}N(t)}$. It denotes the time
elapsed from the last transition of the system. The relevance of this
process in the disability insurance modeling has been described in
\citep{daguma09}.

To characterize the probabilistic evolution of the system we
introduce the following transition probability function:

\begin{defin}
The age-indexed semi-Markov transition probability function with
initial and final backward is the matrix-valued function
\[
^{a+s-u}\boldsymbol{\varPhi}\bigl(u,s;u',t\bigr)= \bigl(
\,^{a+s-u}\phi _{ij}\bigl(u,s;u',t\bigr) \bigr), \quad i,j\in E,\,\, u,s,u',t\in\mathbb{N},
\]
whose generic element $^{a+s-u}\phi_{ij}(u,s;u',t)$ expresses the probability
%
%e3 ###
\begin{equation}
\label{prob} \mathbb{P}\bigl[\,^{a}Z(t)\!=\!j, B(t)\!=
\!u' \bigm\vert \,^{a}Z(s)\!=\!i, B(s)\!=\!u, A_{\,^{a}N(s)}\!=
\!a+T_{\,^{a}N(s)}\bigr].
\end{equation}
\end{defin}
In disability insurance the probability (\ref{prob}) can be
interpreted as the probability that an insured will be at time $t$
%with
in a disability of degree $j$
%of
and duration $u'$ given that at time $s$ she/he was
%with
in a disability of degree $i$ %of
and duration $u$ and of age $a+s$.

\begin{prop}
The age-indexed semi-Markov transition probability function with
initial and final backward satisfy the following recursive system of
equations
%
%e4 ###
\begin{align}
^{a+s-u}\phi_{ij}
\bigl(u,s;u',t\bigr)&=1_{\{i=j\}}1_{\{u'=t-s+u\}}
\frac
{^{a+s-u}\overline{H}_{i}(s-u;t)}{^{a+s-u}\overline{H}_{i}(s-u;s)}\notag
\\[-2pt]
& \quad +\sum_{k\in E}\sum_{\theta=s+1}^{t-u'}
\frac
{^{a+s-u}q_{ik}(s-u;\theta)}{^{a+s-u}\overline{H}_{i}(s-u;s)}\,\cdot \,^{a+\theta}\phi_{kj}\bigl(0,
\theta;u',t\bigr)\xch{,}{.}\label{numero}
\end{align}
where
%
%e5 ###
\begin{align}
^{a+s}q_{ij}(s;t)&=\mathbb{P}[J_{n+1}=j, T_{n+1}= t \mid J_{n}=i, T_{n}=s, A_{n}=a+s]\notag\\[-1pt]
& =\lleft\{ %
\begin{matrix}
^{a+s}Q_{ij}(s;t)-\,^{a+s}Q_{ij}(s;t-1) & \textrm{if} \ t > s, \\
0 & \textrm{if} \  \xch{t = s.}{t = s}
\end{matrix} %
\rright.\label{kernel}
\end{align}
\end{prop}
\endgroup
\begin{proof}
Let us denote by $\mathbb{P}_{(i,s-u,a+s-u)}(\cdot)$ the probability measure
\[
\mathbb{P}\bigl(\cdot\bigm\vert\,^{a}Z(s)\!=\!i, T_{\,^{a}N(s)}=s-u,
A_{\,
^{a}N(s)}\!=\!a+s-u\bigr),
\]
and by $\mathbb{P}_{(i,s-u,a+s-u, >s)}(\cdot)$ the probability measure
\[
\mathbb{P}\bigl(\cdot\bigm\vert\,^{a}Z(s)\!=\!i, T_{\,^{a}N(s)}=s-u,
A_{\,
^{a}N(s)}\!=\!a+s-u, T_{\,^{a}N(s)+1}>s\bigr).
\]

Observe that the information set $\{^{a}Z(s)\!=\!i, B(s)\!=\!u,
A_{\,^{a}N(s)}\!=\!a+T_{\,^{a}N(s)}\}$ is equivalent to $\{^{a}Z(s)\!=\!
i, T_{\,^{a}N(s)}=s-u, T_{\,^{a}N(s)+1}>s, A_{\,^{a}N(s)}\!=\!a+s-u\}$,
so that the age-indexed semi-Markov transition probability function can
be denoted by
%
%e6 ###
\begin{align}
^{a+s-u}\phi_{ij}\bigl(u,s;u',t\bigr)&=\mathbb{P}_{(i,s-u,a+s-u,>s)}\xch{\bigl[\, ^{a}Z(t)\!=\!j, B(t) = u'\bigr]}{\bigl[\, ^{a}Z(t)\!=\!j, B(t) = u'\bigr].}\notag\\
&= \mathbb{P}_{(i,s-u,a+s-u,>s)}\bigl[\,^{a}Z(t)\!=\!j, T_{\,^{a}N(t)}=t-u', T_{\,^{a}N(s)+1}>t\bigr]\notag\\
&\quad  {+}\, \mathbb{P}_{(i,s-u,a+s-u,>s)}\bigl[\,^{a}Z(t)\!=\!j,\! T_{\,^{a}N(t)}\,{=}\,t\,{-}\,u', T_{\,^{a}N(s)+1}\,{\leq}\, \xch{t\bigr].}{t\bigr]} \label{trans}
\end{align}
The first
%addendum
summand of (\ref{trans}) can be represented as follows:
\begin{align*}
& \frac{\mathbb{P}_{(i,s-u,a+s-u,>s)}[\,T_{\,^{a}N(s)+1}>t , ^{a}Z(t)\!=\!j, T_{\,^{a}N(t)}=t-u' ]}{\mathbb{P}_{(i,s-u,a+s-u,>s)}[T_{\,^{a}N(s)+1}>s]}\\
& \quad = \frac{1}{\mathbb{P}_{(i,s-u,a+s-u,>s)}[T_{\,^{a}N(s)+1}>s]}\\
& \qquad \cdot \bigl(\mathbb{P}_{(i,s-u,a+s-u,>s)}\bigl[\,T_{\,^{a}N(s)+1}>t , ^{a}Z(t)\!=\!j, T_{\,^{a}N(t)}=t-u'\bigr]\\
& \qquad \cdot\mathbb{P}_{(i,s-u,a+s-u,>s)}\bigl[\,T_{\,^{a}N(t)}=t-u'\bigr]\\
& \qquad \cdot\mathbb{P}\bigl[T_{\,^{a}N(s)+1}>t \bigm\vert \, ^{a}Z(s)\!=\!i, T_{\, ^{a}N(s)}=s-u, A_{\,^{a}N(s)}\!=\!a+s-u\bigr] \bigr)\\
& \quad = \frac{1}{^{a+s-u}\overline{H}_{i}(s-u;s)}\cdot \bigl(1_{\{i=j\}} \cdot1_{\{u'=t-s+u\}}\cdot^{a+s-u}\overline{H}_{i}(s-u;t) \bigr).
\end{align*}
The second %addendum
summand of (\ref{trans}) can be represented as follows:
\begin{align*}
&\frac{\mathbb{P}_{(i,s-u,a+s-u)}[\,^{a}Z(t)=j, T_{\,^{a}N(t)}=t-u', s<T_{\,^{a}N(s)+1}\leq t]}{\mathbb{P}_{(i,s-u,a+s-u,>s)}[T_{\, ^{a}N(s)+1}>s]}\\
& \quad =\frac{1}{^{a+s-u}\overline{H}_{i}(s-u;s)}\sum_{k\in E}\sum_{\theta=s+1}^{t-u'}\mathbb{P}_{(i,s-u,a+s-u)}\bigl[\,^{a}Z(t)=j, T_{\,^{a}N(t)}=t-u',\\
& \quad \quad \quad  J_{\,^{a}N(s)+1}=k, T_{\,^{a}N(s)+1}= \theta\bigr]\\
& \quad =\frac{1}{^{a+s-u}\overline{H}_{i}(s-u;s)}\\
& \quad \quad \!\!\cdot\!\! \sum_{k\in E}\sum_{\theta=s+1}^{t-u'}\!\!\!\mathbb {P}_{(i,s-u,a+s-u)}\bigl[\,^{a}Z(t)\!=\!j,\! T_{\,^{a}N(t)}\,{=}\,t\,{-}\,u' \bigm\vert J_{\,^{a}N(s)+1}\!=\!k, T_{\,^{a}N(s)+1}\!=\! \theta\bigr]\\
& \quad \quad \!\!\cdot\mathbb{P}_{(i,s-u,a+s-u)}[\, J_{\,^{a}N(s)+1}=k, T_{\,^{a}N(s)+1}=\theta]\\
& \quad = \sum_{k\in E}\sum_{\theta=s+1}^{t-u'}\frac{^{a+s-u}q_{ik}(s-u;\theta)}{^{a+s-u}\overline{H}_{i}(s-u;s)}\,\cdot \,^{a+\theta}\phi_{kj}\bigl(0,\theta;u',t\bigr).
\end{align*}
The last equality is obtained using the assumption (\ref{due}) on the
Markovianity of the triple $(J_{n}, T_{n}, A_{n})$
with respect to transition times $T_{n}$ and the definition of the
age-indexed semi-Markov kernel given in formula (\ref{kernel}).
\end{proof}

The above-presented transition probabilities generalize the
corresponding transition probabilities with initial backward derived in
\citep{dajama11} by including the dependence on the final backward.
Moreover they generalize the transition probabilities with initial and
final backward given in \citep{daguma09} by including the dependence
on the age-index process.

In the sequel of the paper we need to consider survival functions for
our age-indexed model.
To this end we introduce the hitting time of state $D$ (death of the
policyholder) given the occupancy of state $i$ at time $s$ with age $a
+ s$ and duration in the state equal to $u$:
\begin{equation*}
^{a+s-u}T_{i,D}(u,s) := \inf\bigl\{t>s:\, ^{a}Z(t)=D \bigm\vert \,^{a}Z(s)=i, B(s)=u\bigr\}.
\end{equation*}

\begin{defin}
The survival function of the age-indexed semi-Markov chain is the
vector valued function
$^{a+s-u}\mathbf{S}(u,s;t)= (\,^{a+s-u}S_{i}(u,s;t) )$, $i\in
E$, $u,s,t\in\mathbb{N}$ with generic element given by:
%
%e7 ###
\begin{equation}
^{a+s-u}S_{i}(u,s;t):=\mathbb{P}\bigl[\,^{a}T_{i,D}(u,s)>t
\bigr].
\end{equation}
\end{defin}
It denotes the probability to not enter state $D$ in the time interval
$(s,t]$ given the occupancy of state $i$ at time $s$ being aged $a+s$
with entrance in this state with last transition $u$ periods before.
This function can be calculated using the following relation:
\begin{equation*}
^{a+s-u}S_{i}(u,s;t)=\sum_{j\neq D}\sum
_{u'=0}^{t-s+u}\,^{a+s-u}\phi
_{ij}\bigl(u,s;u',t\bigr).
\end{equation*}

It is simple to note that
%
%e8 ###
\begin{align}
\mathbb{P}\bigl[\,^{a+s-u}T_{i,D}(u,s)=t
\bigr]&=\,^{a+s-u}S_{i}(u,s;t-1)-\, ^{a+s-u}S_{i}(u,s;t)\notag\\
& =:\varDelta^{a+s-u}S_{i}(u,s;t-1).
\end{align}

%s3 ###
\section{The conversion option in life insurance}\label{sec3}

\begin{figure}[htpb]
\includegraphics{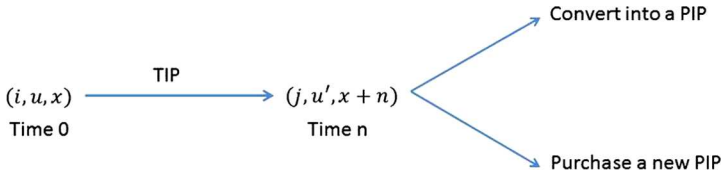}
\caption{A conversion option diagram}
\label{conversion}
\end{figure}

Let us consider the general situation where a female insured aged $x$
at the initial time $0$ with
%an
a health state $i\in E$ buys an $n$-year term insurance policy (TIP).
When the policy is almost due, if she is still alive
she decides to extend the policy for the rest of her life. The
extension can be done by converting the initial TIP into a PIP or
buying a new PIP. In \xch{Figure}{figure} \ref{conversion} we report a diagram
that summarizes the time schedule of a conversion option contract. It
should be remarked that at time $n$, the decision to convert the TIP
into a PIP or to purchase a new PIP should be taken considering the new
health state of the policyholder ($^{a}Z(n)$), the duration in this
state ($B(n)$) and the age ($x+n$).
%
%f1 ###

The valuation of the conversion option
needs the study of two kinds of contracts involved here: the TIP and
the PIP contracts.

%s3.1 ###
\subsection{Temporary insurance policy contract}

Term insurance policies provide coverage for a limited time ($n$ years)
and gives to the policyholder a benefit in case of death. In this paper
without loss of generality we assume that the benefit is set to 1 Euro.
The possession of this coverage is subordinated to the payment, by the
policyholder, of an yearly premium until the occurrence of the death
event or the expiry of the contract whichever occur before.

%Relatively to
For the TIP contract, let us introduce the random variable (r.v.) {\em
conditional Present Value of Death Benefit} denoted by
$(\mathit{PVDB})_{i,u,x}$. It takes value $\delta^{s}$ when the death of the
policyholder occurs at any time $s\leq n$. Given the initial conditions
$\{^{a}Z(0)=i, B(0)=u, A(0)=x\}$, the death event may
%occurs
occur at time $s$ with probability $^{x}S_{i}(u,0;s-1)-\,
^{x}S_{i}(u,0;s)$, then it results
%that
in
\begin{equation*}
\begin{aligned} \mathcal{A}_{i,u}(x,0,n):=&\,\mathbb{E}
\bigl[(\mathit{PVDB})_{i,u,x}\bigr]=\sum_{s=1}^{n}
\mathbb{P}\bigl[^{x}T_{i,D}(u,0)=s\bigr]\cdot1\cdot
\delta^{s}
\\
=&\,\sum_{s=1}^{n} \varDelta^{x}S_{i}(u,0;s-1)
\delta^{s}. \end{aligned} %
\end{equation*}
Let us introduce the r.v. {\em conditional Present Value of Unitary
Premiums} denoted by $(\mathit{PVUP})_{i,u,x}$. Since premiums are paid in the
due case, the r.v. $(\mathit{PVUP})_{i,u,x}$ takes value $\sum_{r=0}^{s-1}\delta^{r}$ when the death of the policyholder occurs at
time $s\leq n-1$ and value $\sum_{r=0}^{n}\delta^{r}$ if she will
survive time $n$.

Let us denote by $p_{i,u}(x,0)$ the annual premium \xch{for an}{for a} $n$-TIP with $1$
Euro payable at the year of death of an insured
of age $x$, in health state $i$ obtained $u$ years before. Then the
r.v. {\em conditional Present Value of Premiums} denoted by
$(\mathit{PVP})_{i,u,x}$ is simply defined by
%
%e9 ###
\begin{align}
& (\mathit{PVP})_{i,u,x}:=p_{i,u}(x,0)\cdot(\mathit{PVUP})_{i,u,x}, \quad  \textrm{for}\, i\neq \xch{D,}{D}\notag\\
& (\mathit{PVP})_{i,u,x} := 0, \quad  \textrm{for}\ \xch{i=D,}{i=D.}
\end{align}
and then it results %that :
in
\begin{align*}
 \mathcal{P}_{i,u}(x,0,n)&:=\mathbb{E}\bigl[(\mathit{PVP})_{i,u,x}\bigr]\\
&\,=\sum_{s=1}^{n-1} \Biggl(p_{i,u}(x,0)\sum_{r=0}^{s-1}\delta^{r} \Biggr)\varDelta\,^{x}S_{i}(u,0;s-1)\\
&\,\quad + \Biggl(p_{i,u}(x,0)\sum_{r=1}^{n}\delta ^{r} \Biggr)\,^{x}S_{i}(u,0;n).
\end{align*}

Furthermore if we assume that premiums are fixed according to
the equivalence principle, i.e. in a way such that the actuarial
present value of premiums should be equal to the actuarial present
value of benefits (see e.g. \citep{hapi99}), then we have that:
\begingroup
\abovedisplayskip=4.5pt
\belowdisplayskip=4.5pt
\begin{equation*}
\mathcal{A}_{i,u}(x,0,n)= \mathcal{P}_{i,u}\xch{(x,0,n),}{(x,0,n)}
\end{equation*}
from which we recover the fair premium
%
%e10 ###
\begin{equation}
p_{i,u}(x,0)=\frac{\sum_{s=1}^{n} \varDelta\,^{x}S_{i}(u,0;s-1) \delta
^{s}}{\sum_{s=1}^{n-1}\sum_{r=1}^{s-1}\delta^{r} \varDelta\,
^{x}S_{i}(u,0;s-1)+\sum_{r=1}^{n}\delta^{r}\,^{x}S_{i}(u,0;n)}.
\end{equation}

%s3.2 ###
\subsection{Permanent insurance policy}

Permanent insurance policies provide coverage for an unlimited time
horizon and gives to the policyholder a benefit of 1 Euro in case of
death. The possession of this coverage is subordinated to the payment,
by the policyholder, of an yearly premium until the occurrence of the
death event.

Relatively to the PIP contract let us introduce the r.v. {\em
conditional Present Value of Death Benefits} denoted by $(\widetilde
{\mathit{PVDB}})_{i,u,x}$. It takes value $\delta^{s}$ when the death of the
policyholder occurs at time $s\in\mathbb{N}$. In analogy with the TIP
case it results
%that
in
\begin{equation*}
\begin{aligned} \tilde{\mathcal{A}}_{i,u}(x,0)&:=\mathbb{E}
\bigl[(\widetilde {\mathit{PVDB}})_{i,u,x}\bigr]=\sum_{s=1}^{\infty}
\mathbb {P}\bigl[^{x}T_{i,D}(u,0)=s\bigr]\cdot1\cdot
\delta^{s}
\\
& \,=\sum_{s=1}^{\infty} \varDelta^{x}S_{i}(u,0;s-1)
\delta^{s}. \end{aligned} %
\end{equation*}
Let us introduce the r.v. {\em conditional Present Value of Unitary
Premiums} denoted by $(\widetilde{\mathit{PVUP}})_{i,u,x}$. Premiums are paid
until the occurrence of the death of the policyholder, formally the
r.v. $(\widetilde{\mathit{PVUP}})_{i,u,x}$ assumes value $\sum_{r=1}^{s-1}\delta^{r}$ when the death of the policyholder occurs at
time $s\in\mathbb{N}$.

Let us denote by $\tilde{p}_{i,u}(x,0)$ the annual premium for a PIP
with $1$ Euro payable at the year of death of an insured
of age $x$, in health state $i$ obtained $u$ years before. Then the
r.v. {\em conditional Present Value of Premiums} denoted by
$(\widetilde{\mathit{PVP}})_{i,u,x}$ is simply defined by
%
%e11 ###
\begin{align}
& (\widetilde{\mathit{PVP}})_{i,u,x}:=\tilde{p}_{i,u}(x,0)\cdot\xch{(\widetilde {\mathit{PVUP}})_{i,u,x},}{(\widetilde {\mathit{PVUP}})_{i,u,x}}\quad \textrm{for}\ i\neq D,\notag\\
& (\widetilde{\mathit{PVP}})_{i,u,x} := 0, \quad \textrm{for}\ \xch{i=D,}{i=D.}
\end{align}
and then it results %that :
in
\begin{equation*}
\tilde{\mathcal{P}}_{i,u}(x,0):=\mathbb{E}\bigl[(\widetilde
{\mathit{PVP}})_{i,u,x}\bigr]=\sum_{s=1}^{\infty}
\tilde{p}_{i,u}(x,0)\sum_{r=1}^{s-1}
\delta^{r}\varDelta\,^{x}S_{i}(u,0;s-1).
\end{equation*}

Furthermore if we assume that premiums are fixed according to
the equivalence principle we have that:
\begin{equation*}
\tilde{\mathcal{A}}_{i,u}(x,0)= \tilde{\mathcal{P}}_{i,u}(x,0),
\end{equation*}
\endgroup
from which we recover the fair premium
%
%e12 ###
\begin{equation}
\tilde{p}_{i,u}(x,0)=\frac{\sum_{s=1}^{\infty}\delta^{s} \varDelta\,
^{x}S_{i}(u,0;s-1)}{\sum_{s=1}^{\infty}\sum_{r=1}^{s-1}\delta
^{r}\varDelta\,^{x}S_{i}(u,0;s-1)}.
\end{equation}

%s3.3 ###
\subsection{Valuation of the conversion option}

In this subsection we develop the valuation procedure for conversion
options when survival probability functions are derived from a
multi-state model of the policyholder's health. The valuation makes use
of the random variables introduced for describing the TIP and PIP
contracts and what we called {\em exercise set} of the option. The
introduction of the exercise set is a prerogative of our model and was
not present in earlier studies on conversion options. We remember that
the policyholder possesses a TIP issued at time zero with maturity $n$
and at time $n$ should decide to prolong the insurance coverage either
by means of converting the TIP into a PIP or purchasing a new PIP.

We define the r.v. {\em conditional Conversion Gain} as
%
%e13 ###
\begin{equation}
(\mathit{CG})_{i,u,x}= \bigl[(\mathit{PVDB})_{i,u,x} \bigr]- \bigl[(\mathit{PVP})_{i,u,x}
\bigm\vert\textrm {conversion} \bigr],
\end{equation}
where $ [(\mathit{PVDB})_{i,u,x} ]$ is the r.v. denoting the present
value of death benefits and\break $ [(\mathit{PVP})_{i,u,x}\mid\textrm
{conversion} ]$ is the r.v. describing the present value of
premiums when the policyholder possesses an option to convert the
original TIP into a PIP before the expiry of the TIP.

They are both conditional on the information set $\{^{a}Z(0)=i,
B(0)=u,\break
A(0)=x\}$ describing the initial health conditions of the policyholder
at the inception time zero. The formal definition of the r.v. $
[(\mathit{PVP})_{i,u,x}\mid\textrm{conversion} ]$ is given in
Definition~\ref{def4} below.

Similarly it is possible to define the r.v. {\em conditional No
Conversion Gain} as
%
%e14 ###
\begin{equation}
(\mathit{NCG})_{i,u,x}= \bigl[(\mathit{PVDB})_{i,u,x} \bigr]- \bigl[(\mathit{PVP})_{i,u,x}
\bigm\vert \textrm{no conversion} \bigr],
\end{equation}
where $ [(\mathit{PVP})_{i,u,x}\mid\textrm{no conversion} ]$ is the
r.v. denoting the present value of premiums when the policyholder does
not possess an option to convert the original TIP into a PIP and then
must purchase a new PIP at time $n$
if she wants to extend the insurance protection. The formal definition
of the r.v. $ [(\mathit{PVP})_{i,u,x}\mid\textrm{no conversion} ]$
is given in Definition~\ref{def3} below.

The difference between the Conversion Gain and the No Conversion Gain
define the r.v.
{\em conditional Net Gain}, i.e.:
%
%e15 ###
\begin{equation}
(G)_{i,u,x}=(\mathit{CG})_{i,u,x}-(\mathit{NCG})_{i,u,x},
\end{equation}
and its expected value is called conditional {\em Value of the
Conversion Option}, i.e.:
%
%e16 ###
\begin{equation}
(\mathit{VCO})_{i,u,x}=\mathbb{E}\bigl[(G)_{i,u,x}\bigr].
\end{equation}
It is simple to realize that
%
%e17 ###
\begin{equation}
\label{vco} (\mathit{VCO})_{i,u,x}=\mathbb{E} \bigl[(\mathit{PVP})_{i,u,x}\bigm\vert
\textrm{no conversion} \bigr]-\mathbb{E} \bigl[(\mathit{PVP})_{i,u,x}\bigm\vert\textrm
{conversion} \bigr].
\end{equation}

Therefore, we need to calculate the expectations on the right hand side
of \xch{Eq.}{equation}~(\ref{vco}).
To do this we proceed first %before
to the formal definition of the two random variables involved in the
computation. This requires the introduction of some auxiliary concepts.

Let us consider a time $n\in\mathbb{N}$, then the triple $(i,u,x)$ is
called an {\em n-scenario} if $^{a}Z(n)=i$, $B(n)=u$, $A_{N(n)}=x-n+u$.\vadjust{\goodbreak}

We say that the 0-scenario $(i,u,x)$ is state-unchanged at time $n$ if
the $n$-scenario will be $(i,u,x+n)$.

Two state-unchanged scenarios share the same health state and duration
in this state but are characterized by different ages of the policyholder.

The conditional {\em cash Value} is defined by
%
%e18 ###
\begin{equation}
V_{i,u}(x+n,n):=\bigl[\tilde{p}_{i,u}(x+n,n)-p_{i,u}(x+n,0)
\bigr]\cdot \widetilde{\mathit{PVUP}}_{i,u,x}\cdot\delta^{n}.
\end{equation}

The expectation of the cash value is the quantity the
policyholder has to pay at the time of conversion to the insurance company:
%
%e19 ###
\begin{equation}
\begin{aligned}
\mathcal{V}_{i,u}(x+n,n)&:=\mathbb{E}\bigl[V_{i,u}(x+n,n)\bigr]\\
&\,= \bigl[\tilde{p}_{i,u}(x+n,n)-p_{i,u}(x+n,0)\bigr]\cdot\sum_{h=n+1}^{\infty}\delta^{h}\,\varDelta\,^{x+n}S_{i}(u,n;h-1).
\end{aligned}
\end{equation}

The quantity $\mathcal{V}_{i,u}(x+n,n)$ expresses the gain the
policyholder expect to realize buying the conversion option under the
hypothesis of an unchanged $n$-scenario. This quantity is greater or
equal than zero because
\[
\tilde{p}_{i,u}(x+n,n)\geq p_{i,u}(x+n,0),
\]
that is, the premiums for a PIP are greater than the corresponding
premium for a TIP given the same $n$-scenario $(i,u,x+n)$.

In analogy with the financial options, we can define a set where it is
convenient to exercise the conversion option. This is a prerogative of
the adopted multi-state model because in the paper
%by
\citep{su10}, if the insured person was still alive at the conversion
time it was always convenient to prolong the coverage by exercising the
option. However, in our more general framework, this is not the case,
because given the initial 0-scenario $(i,u,x)$ it is possible after $n$
years that the insured person improves considerably the health state
and the prospective
%of a prolonged life expectancy.
expectation of a prolonged life.
This has been observed in the evolution of several diseases like HIV
infection, see e.g.~\citep{dadijama11}.\looseness=1

Given the 0-scenario $(i,u,x)$, we define the {\em exercise set} as
%
%e20 ###
\begin{align}
C_{i,u}(x,n):=& \bigl\{\bigl(j,u'\bigr)\in E\times\mathbb{N} :\notag\\
& \mathbb{E}\bigl[p_{i,u}(x,0)\cdot\widetilde {\mathit{PVUP}}_{i,u,x}+V_{i,u}(x+n,n)\bigr]\leq\tilde{\mathcal {P}}_{j,u'}(x+n,n) \bigr\}.
\end{align}

The set $C_{i,u}(x,n)$ comprehends all couples of health states and
durations where it is convenient for the policyholder to exercise the
conversion option. Indeed, if the expected payment to face by
converting the option $\mathbb{E}[p_{i,u}(x,0)\cdot\widetilde
{\mathit{PVUP}}_{i,u,x}+V_{i,u}(x+n,n)]$ is smaller than the expected present
value of premiums to be paid for a new PIP in the new $n$-scenario
$(j,u',x+n)$ it is convenient to convert the option because with an
inferior cost the policyholder guarantees to
%itself
herself the same insurance protection. Therefore, if $(j,u')\in
C_{i,u}(x,n)$ the policyholder will convert the option; on the
contrary, if $(j,u')\in C_{i,u}^{c}(x,n)$ the policyholder will not
convert the option.\looseness=1

Now we are in the position to define the random variables
\[
\bigl[(\mathit{PVP})_{i,u,x}\bigm\vert\textrm{no conversion}\bigr], \,\,\,\,
\bigl[(\mathit{PVP})_{i,u,x}\bigm\vert\textrm{conversion}\bigr].
\]

\begin{defin}\label{def3}
The r.v. $[(\mathit{PVP})_{i,u,x}\mid\textrm{no conversion}]$ is defined by
the following relation:
%
%e21 ###
\begin{equation}
\bigl[(\mathit{PVP})_{i,u,x}\bigm\vert\textrm{no conversion}\bigr]:=(\mathit{PVP})_{i,u,x}+(
\widetilde {\mathit{PVP}})_{\,^{a}Z(n),B(n),A(n)}\cdot\delta^{n}.
\end{equation}
\end{defin}

Then, the conditional present value of premiums given no
conversion is equal to the conditional present value of premiums from
the TIP contract plus the conditional present value of premiums of the
subsequent PIP calculated under the $n$-scenario $(^{a}Z(n),B(n),A(n))$
and discounted at time zero.

It is possible to calculate its expectation that is given here below:
%
%e22 ###
\begin{equation}
\begin{aligned} \mathbb{E}[\mathit{PVP}\mid\textrm{no conversion}]&=\mathcal
{P}_{i,u}(x,0,n)
\\
& \quad + \sum_{j\in E} \sum_{u'\geq0}
\,^{x}\phi_{ij}\bigl(u,0;u',n\bigr)\cdot
\delta^{n}\cdot\tilde{\mathcal{P}}_{j,u'}(x+n,n). \end{aligned}
\end{equation}
\begin{defin}\label{def4}
The r.v. $[(\mathit{PVP})_{i,u,x}\mid\textrm{conversion}]$ is defined by the
following relation:
%
%e23 ###
\begin{align}
\bigl[(\mathit{PVP})_{i,u,x}\bigm\vert\textrm{conversion}\bigr]&:=(\mathit{PVP})_{i,u,x}\notag\\
&\quad  +\delta^{n} (\widetilde{\mathit{PVP}})_{\,^{a}Z(n),B(n),A(n)}\cdot1_{\{(^{a}Z(n),B(n))\in C_{i,u}^{c}(x,n)\}}\notag\\
&\quad  + \delta^{n} \bigl[\bigl(p_{i,u}(x,0)\widetilde{\mathit{PVUP}}\bigr) + V_{i,u}(x+n,n)\bigr]\notag\\
&\quad \cdot 1_{\{(^{a}Z(n),B(n))\in C_{i,u}(x,n)\}}.\label{pvpconv}
\end{align}
\end{defin}

Then, the conditional present value of premiums given the
possibility to convert is equal to the conditional present value of
premiums from the TIP contract plus the conditional present value of
premiums from the PIP calculated under the $n$-scenario
$(^{a}Z(n),B(n),A(n))$ and discounted at time zero if this scenario
does not belong to the exercise set plus the expected payment to face
by converting the option if the $n$-scenario belongs to the exercise
set.

It is possible to calculate the expectation of (\ref{pvpconv})
that is given here below:
%
%\begin{equation}
%e24 ###
\begin{align}
\mathbb{E}[\mathit{PVP}\mid\textrm{conversion}]&=\mathcal{P}_{i,u}(x,0,n)\notag\\
& \quad + \sum_{(j,u')\in C_{i,u}^{c}(x,n)}\,^{x}\phi_{ij}\bigl(u,0;u',n\bigr)\cdot \delta^{n}\cdot\tilde{\mathcal{P}}_{j,u'}(x+n,n)\notag\\
& \quad + \sum_{(j,u')\in C_{i,u}(x,n)}\,^{x}\phi_{ij}\bigl(u,0;u',n\bigr)\cdot \delta^{n}\cdot\Biggl[V_{i,u}(x+n,n)\notag\\
& \quad +\sum_{h=n+1}^{\infty}p_{i,u}(x,0)\sum_{r=n+1}^{h}\delta ^{r}\varDelta\,^{x+n}S_{j}\bigl(u',n;h\bigr) \xch{\Biggr].}{\Biggr]}\label{pvpnoconv}
\end{align}
%
%\end{equation}

Now we are in the position of computing the value of the
conversion option by substituting \xch{Eqs}{equations} (\ref{pvpconv}) and (\ref
{pvpnoconv}) in Formula (\ref{vco}). Some algebra gives the following
representation:
\begin{align*}
(\mathit{VCO})_{i,u,x}&=\sum_{(j,u')\in C_{i,u}(x,n)}\!\!\!\!\!\!\!\,^{x}\phi _{ij}\bigl(u,0;u',n\bigr)\delta^{n}\cdot \Biggl[\tilde{\mathcal{P}}_{j,u'}(x+n,n)- V_{i,u}(x+n,n)\\
&\quad  -\sum_{h=n+1}^{\infty}p_{i,u}(x,0)\sum_{r=n+1}^{h}\delta^{r} \bigl(\,^{x+n}S_{j}\bigl(u',n;h\bigr){-}\,^{x+n}S_{j}\bigl(u',n;h+1\bigr) \bigr) \!\Biggr],
\end{align*}
from which we realize that $\mathit{VCO}\geq0$ because on the exercise set
$C_{i,u}(x,n)$ the term within square brackets is nonnegative.

We would like to remark that the value of the conversion option is
nonnegative unless the exercise set is empty. Moreover the value does
depend on the
%dynamic
dynamics of the health state of the policyholder and therefore, in our
model, it is sensitive to the duration of permanence in the health
state, to the chronological time and to the age of the policyholder.

%s4 ###
\section{Conclusions}\label{sec4}
The valuation of conversion options in life insurance is an important
subject in modern actuarial mathematics.

This study accomplished several goals. First, we proposed a general
multistate model that can reproduce important aspects in the modeling
of life insurance contracts and we calculated transition probability
function for the model. Second, we defined the main variables necessary
to the description of the contract and we calculated the value of the
conversion option in a very general framework. As particular cases we
obtain formulas for the valuation of temporary insurance policy and
permanent insurance policy that are embedded in the conversion option contract.

This paper leaves several points opened. First of all the application
to real data of the model is by far the most urgent task to be
accomplished. This task can be accomplished once a reliable dataset is
obtained and adequate computer programmes are built. Then, the
possibility to extend the results to more complex models is also relevant,
in this light
a possible extension to subordinated semi-Markov chains
is worth mentioning.

\bibliographystyle{vmsta-mathphys}
%\bibliography{biblio}

\begin{thebibliography}{18}

%b1 ###
\bibitem{da11}
\begin{barticle}
\bauthor{\bsnm{D'Amico}, \binits{G.}}:
\batitle{Age-usage semi-Markov models}.
\bjtitle{Appl. Math. Model.}
\bvolume{35},
\bfpage{4354}--\blpage{4366}
(\byear{2011}).
\bid{doi={10.1016/j.apm.2011.03.006}, mr={2801959}}
\end{barticle}
%
\OrigBibText
\begin{barticle}
\bauthor{\bsnm{D'Amico}, \binits{G.}}:
\batitle{Age-usage semi-markov models}.
\bjtitle{Applied Mathematical Modelling}
\bvolume{35},
\bfpage{4354}--\blpage{4366}
(\byear{2011})
\end{barticle}
\endOrigBibText
\bptok{structpyb}%
\endbibitem

%b2 ###
\bibitem{dape11}
\begin{barticle}
\bauthor{\bsnm{D'Amico}, \binits{G.}},
\bauthor{\bsnm{Petroni}, \binits{F.}}:
\batitle{A semi-Markov model with memory for price changes}.
\bjtitle{J. Stat. Mech. Theory Exp.},
\bfpage{P12009}
(\byear{2011})
\end{barticle}
%
\OrigBibText
\begin{botherref}
\oauthor{\bsnm{D'Amico}, \binits{G.}},
\oauthor{\bsnm{Petroni}, \binits{F.}}:
A semi-markov model with memory for price changes.
J. Stat. Mech. Theory Exp.
\textbf{P12009}
(2011)
\end{botherref}
\endOrigBibText
\bptok{structpyb}%
\endbibitem

%b3 ###
\bibitem{dape12}
\begin{barticle}
\bauthor{\bsnm{D'Amico}, \binits{G.}},
\bauthor{\bsnm{Petroni}, \binits{F.}}:
\batitle{Weighted-indexed semi-Markov models for modeling financial returns}.
\bjtitle{J. Stat. Mech. Theory Exp.},
\bfpage{P07015}
(\byear{2011})
\end{barticle}
%
\OrigBibText
\begin{botherref}
\oauthor{\bsnm{D'Amico}, \binits{G.}},
\oauthor{\bsnm{Petroni}, \binits{F.}}:
Weighted-indexed semi-markov models for modeling financial returns.
J. Stat. Mech. Theory Exp.
\textbf{P07015}
(2011)
\end{botherref}
\endOrigBibText
\bptok{structpyb}%
\endbibitem

%b4 ###
\bibitem{daguma09}
\begin{barticle}
\bauthor{\bsnm{D'Amico}, \binits{G.}},
\bauthor{\bsnm{Guillen}, \binits{M.}},
\bauthor{\bsnm{Manca}, \binits{R.}}:
\batitle{Full backward non-homogeneous semi-Markov processes for
disability
 insurance models: A Catalunya real data application}.
\bjtitle{Insur. Math. Econ.}
\bvolume{45},
\bfpage{173}--\blpage{179}
(\byear{2009}).
\bid{doi={10.1016/j.insmatheco.\\2009.05.010}, mr={2583371}}
\end{barticle}
%
\OrigBibText
\begin{barticle}
\bauthor{\bsnm{D'Amico}, \binits{G.}},
\bauthor{\bsnm{Guillen}, \binits{M.}},
\bauthor{\bsnm{Manca}, \binits{R.}}:
\batitle{Full backward non-homogeneous semi-markov processes for
disability
 insurance models: A catalunya real data application}.
\bjtitle{Insurance: Mathematics and Economics}
\bvolume{45},
\bfpage{173}--\blpage{179}
(\byear{2009})
\end{barticle}
\endOrigBibText
\bptok{structpyb}%
\endbibitem

%b5 ###
\bibitem{daguma13}
\begin{barticle}
\bauthor{\bsnm{D'Amico}, \binits{G.}},
\bauthor{\bsnm{Guillen}, \binits{M.}},
\bauthor{\bsnm{Manca}, \binits{R.}}:
\batitle{Semi-Markov disability insurance models}.
\bjtitle{Commun. Stat., Theory Methods}
\bvolume{42(16)},
\bfpage{2172}--\blpage{2188}
(\byear{2013}).
\bid{doi={10.1080/\\03610926.2012.746982}, mr={3170905}}
\end{barticle}
%
\OrigBibText
\begin{barticle}
\bauthor{\bsnm{D'Amico}, \binits{G.}},
\bauthor{\bsnm{Guillen}, \binits{M.}},
\bauthor{\bsnm{Manca}, \binits{R.}}:
\batitle{Semi-markov disability insurance models}.
\bjtitle{Communications in Statistics - Theory and Methods}
\bvolume{42(16)},
\bfpage{2172}--\blpage{2188}
(\byear{2013})
\end{barticle}
\endOrigBibText
\bptok{structpyb}%
\endbibitem

%b6 ###
\bibitem{dajama11}
\begin{barticle}
\bauthor{\bsnm{D'Amico}, \binits{G.}},
\bauthor{\bsnm{Janssen}, \binits{J.}},
\bauthor{\bsnm{Manca}, \binits{R.}}:
\batitle{Discrete time non-homogeneous semi-Markov reliability
transition
 credit risk models and the default distribution functions}.
\bjtitle{Comput. Econ.}
\bvolume{38},
\bfpage{465}--\blpage{481}
(\byear{2011})
\end{barticle}
%
\OrigBibText
\begin{barticle}
\bauthor{\bsnm{D'Amico}, \binits{G.}},
\bauthor{\bsnm{Janssen}, \binits{J.}},
\bauthor{\bsnm{Manca}, \binits{R.}}:
\batitle{Discrete time non-homogeneous semi-markov reliability
transition
 credit risk models and the default distribution functions}.
\bjtitle{Computational Economics}
\bvolume{38},
\bfpage{465}--\blpage{481}
(\byear{2011})
\end{barticle}
\endOrigBibText
\bptok{structpyb}%
\endbibitem

%b7 ###
\bibitem{dadijama11}
\begin{barticle}
\bauthor{\bsnm{D'Amico}, \binits{G.}},
\bauthor{\bsnm{Di~Biase}, \binits{G.}},
\bauthor{\bsnm{Janssen}, \binits{J.}},
\bauthor{\bsnm{Manca}, \binits{R.}}:
\batitle{HIV evolution: A quantification of the effects due to age and
to
 medical progress}.
\bjtitle{Informatica}
\bvolume{22}(\bissue{1}),
\bfpage{27}--\blpage{42}
(\byear{2011}).
\bid{mr={2885657}}
\end{barticle}
%
\OrigBibText
\begin{barticle}
\bauthor{\bsnm{D'Amico}, \binits{G.}},
\bauthor{\bsnm{Di~Biase}, \binits{G.}},
\bauthor{\bsnm{Janssen}, \binits{J.}},
\bauthor{\bsnm{Manca}, \binits{R.}}:
\batitle{Hiv evolution: A quantification of the effects due to age and
to
 medical progress}.
\bjtitle{Informatica}
\bvolume{22(1)},
\bfpage{27}--\blpage{42}
(\byear{2011})
\end{barticle}
\endOrigBibText
\bptok{structpyb}%
\endbibitem

%b8 ###
\bibitem{hapi99}
\begin{bbook}
\bauthor{\bsnm{Haberman}, \binits{S.}},
\bauthor{\bsnm{Pitacco}, \binits{E.}}:
\bbtitle{Actuarial Models for Disability Insurance}.
\bpublisher{Chapman \& Hall},
\blocation{London}
(\byear{1999}).
\bid{mr={1653961}}
\end{bbook}
%
\OrigBibText
\begin{bbook}
\bauthor{\bsnm{Haberman}, \binits{S.}},
\bauthor{\bsnm{Pitacco}, \binits{E.}}:
\bbtitle{Actuarial Models for Disability Insurance}.
\bpublisher{Chapman \& Hall},
\blocation{London}
(\byear{1999})
\end{bbook}
\endOrigBibText
\bptok{structpyb}%
\endbibitem

%b9 ###
\bibitem{jama97}
\begin{barticle}
\bauthor{\bsnm{Janssen}, \binits{J.}},
\bauthor{\bsnm{Manca}, \binits{R.}}:
\batitle{A realistic non-homogeneous stochastic pension funds model on
scenario
 basis}.
\bjtitle{Scand. Actuar. J.}
\bvolume{2},
\bfpage{113}--\blpage{137}
(\byear{1997})
\end{barticle}
%
\OrigBibText
\begin{barticle}
\bauthor{\bsnm{Janssen}, \binits{J.}},
\bauthor{\bsnm{Manca}, \binits{R.}}:
\batitle{A realistic non-homogeneous stochastic pension funds model on
scenario
 basis}.
\bjtitle{Scandinavian Actuarial Journal}
\bvolume{2},
\bfpage{113}--\blpage{137}
(\byear{1997})
\end{barticle}
\endOrigBibText
\bptok{structpyb}%
\endbibitem

%b10 ###
\bibitem{kwjo06}
\begin{barticle}
\bauthor{\bsnm{Kwon}, \binits{H.S.}},
\bauthor{\bsnm{Jones}, \binits{B.}}:
\batitle{The impact of the determinants of mortality on life insurance
and
 annuities}.
\bjtitle{Insur. Math. Econ.}
\bvolume{38},
\bfpage{271}--\blpage{288}
(\byear{2006}).
\bid{doi={10.1016/\\j.insmatheco.2005.08.007}, mr={2212527}}
\end{barticle}
%
\OrigBibText
\begin{barticle}
\bauthor{\bsnm{Kwon}, \binits{H.S.}},
\bauthor{\bsnm{Jones}, \binits{B.}}:
\batitle{The impact of the determinants of mortality on life insurance
and
 annuities}.
\bjtitle{Insurance: Mathematics and Economics}
\bvolume{38},
\bfpage{271}--\blpage{288}
(\byear{2006})
\end{barticle}
\endOrigBibText
\bptok{structpyb}%
\endbibitem

%b11 ###
\bibitem{kwjo08}
\begin{barticle}
\bauthor{\bsnm{Kwon}, \binits{H.S.}},
\bauthor{\bsnm{Jones}, \binits{B.}}:
\batitle{Applications of a multi-state risk factor/mortality model in
life
 insurance}.
\bjtitle{Insur. Math. Econ.}
\bvolume{43},
\bfpage{394}--\blpage{402}
(\byear{2008}).
\bid{doi={10.1016/\\j.insmatheco.2008.07.004}, mr={2479585}}
\end{barticle}
%
\OrigBibText
\begin{barticle}
\bauthor{\bsnm{Kwon}, \binits{H.S.}},
\bauthor{\bsnm{Jones}, \binits{B.}}:
\batitle{Applications of a multi-state risk factor/mortality model in
life
 insurance}.
\bjtitle{Insurance: Mathematics and Economics}
\bvolume{43},
\bfpage{394}--\blpage{402}
(\byear{2008})
\end{barticle}
\endOrigBibText
\bptok{structpyb}%
\endbibitem

%b12 ###
\bibitem{lili07}
\begin{barticle}
\bauthor{\bsnm{Lin}, \binits{X.S.}},
\bauthor{\bsnm{Liu}, \binits{X.}}:
\batitle{Markov aging process and phase-type law of mortality}.
\bjtitle{N. Am. Actuar. J.}
\bvolume{11},
\bfpage{92}--\blpage{109}
(\byear{2007}).
\bid{doi={10.1080/10920277.2007.10597486}, mr={2413621}}
\end{barticle}
%
\OrigBibText
\begin{barticle}
\bauthor{\bsnm{Lin}, \binits{X.S.}},
\bauthor{\bsnm{Liu}, \binits{X.}}:
\batitle{Markov aging process and phase-type law of mortality}.
\bjtitle{North American Actuarial Journal}
\bvolume{11},
\bfpage{92}--\blpage{109}
(\byear{2007})
\end{barticle}
\endOrigBibText
\bptok{structpyb}%
\endbibitem

%b13 ###
\bibitem{lili12}
\begin{barticle}
\bauthor{\bsnm{Liu}, \binits{X.}},
\bauthor{\bsnm{Lin}, \binits{X.S.}}:
\batitle{A subordinated Markov model for stochastic mortality}.
\bjtitle{Eur. Actuar. J.}
\bvolume{2},
\bfpage{105}--\blpage{127}
(\byear{2012}).
\bid{doi={10.1007/s13385-012-0047-3}, mr={2954471}}
\end{barticle}
%
\OrigBibText
\begin{barticle}
\bauthor{\bsnm{Liu}, \binits{X.}},
\bauthor{\bsnm{Lin}, \binits{X.S.}}:
\batitle{A subordinated markov model for stochastic mortality}.
\bjtitle{European Actuarial Journal}
\bvolume{2},
\bfpage{105}--\blpage{127}
(\byear{2012})
\end{barticle}
\endOrigBibText
\bptok{structpyb}%
\endbibitem

%b14 ###
\bibitem{ma13}
\begin{barticle}
\bauthor{\bsnm{Maegebier}, \binits{A.}}:
\batitle{Valuation and risk assessment of disability insurance using a
discrete
 time trivariate Markov renewal reward process}.
\bjtitle{Insur. Math. Econ.}
\bvolume{53},
\bfpage{802}--\blpage{811}
(\byear{2013}).
\bid{doi={10.1016/j.insmatheco.2013.09.013}, mr={3130475}}
\end{barticle}
%
\OrigBibText
\begin{barticle}
\bauthor{\bsnm{Maegebier}, \binits{A.}}:
\batitle{Valuation and risk assessment of disability insurance using a
discrete
 time trivariate markov renewal reward process}.
\bjtitle{Insurance: Mathematics and Economics}
\bvolume{53},
\bfpage{802}--\blpage{811}
(\byear{2013})
\end{barticle}
\endOrigBibText
\bptok{structpyb}%
\endbibitem

%b15 ###
\bibitem{no08}
\begin{barticle}
\bauthor{\bsnm{Nordahl}, \binits{H.A.}}:
\batitle{Valuation of life insurance surrender and exchange options}.
\bjtitle{Insur. Math. Econ.}
\bvolume{42},
\bfpage{909}--\blpage{919}
(\byear{2008}).
\bid{doi={10.1016/j.insmatheco.2007.\\10.011}, mr={2435361}}
\end{barticle}
%
\OrigBibText
\begin{barticle}
\bauthor{\bsnm{Nordahl}, \binits{H.A.}}:
\batitle{Valuation of life insurance surrender and exchange options}.
\bjtitle{Insurance: Mathematics and Economics}
\bvolume{42},
\bfpage{909}--\blpage{919}
(\byear{2008})
\end{barticle}
\endOrigBibText
\bptok{structpyb}%
\endbibitem

%b16 ###
\bibitem{stmasi07}
\begin{barticle}
\bauthor{\bsnm{Stenberg}, \binits{F.}},
\bauthor{\bsnm{Manca}, \binits{R.}},
\bauthor{\bsnm{Silvestrov}, \binits{D.}}:
\batitle{An algorithmic approach to discrete time non-homogeneous
backward
 semi-Markov reward processes with an application to disability
insurance}.
\bjtitle{Methodol. Comput. Appl. Probab.}
\bvolume{9},
\bfpage{497}--\blpage{519}
(\byear{2007}).
\bid{doi={10.1007/s11009-006-9012-4}, mr={2404740}}
\end{barticle}
%
\OrigBibText
\begin{barticle}
\bauthor{\bsnm{Stenberg}, \binits{F.}},
\bauthor{\bsnm{Manca}, \binits{R.}},
\bauthor{\bsnm{Silvestrov}, \binits{D.}}:
\batitle{An algorithmic approach to discrete time non-homogeneous
backward
 semi-markov reward processes with an application to disability
insurance}.
\bjtitle{Methodology and Computing in Applied Probability}
\bvolume{9},
\bfpage{497}--\blpage{519}
(\byear{2007})
\end{barticle}
\endOrigBibText
\bptok{structpyb}%
\endbibitem

%b17 ###
\bibitem{su10}
\begin{barticle}
\bauthor{\bsnm{Su}, \binits{K.C.}}:
\batitle{The conversion option in life insurance}.
\bjtitle{Insur. Math. Econ.}
\bvolume{46},
\bfpage{437}--\blpage{442}
(\byear{2010}).
\bid{doi={10.1016/j.insmatheco.2009.12.009}, mr={2642520}}
\end{barticle}
%
\OrigBibText
\begin{barticle}
\bauthor{\bsnm{Su}, \binits{K.C.}}:
\batitle{The conversion option in life insurance}.
\bjtitle{Insurance: Mathematics and Economics}
\bvolume{46},
\bfpage{437}--\blpage{442}
(\byear{2010})
\end{barticle}
\endOrigBibText
\bptok{structpyb}%
\endbibitem

%b18 ###
\bibitem{toma91}
\begin{barticle}
\bauthor{\bsnm{Tolley}, \binits{H.D.}},
\bauthor{\bsnm{Manton}, \binits{K.G.}}:
\batitle{Intervention effects among a collection of risks}.
\bjtitle{Trans. Soc. Actuar.}
\bvolume{43},
\bfpage{443}--\blpage{467}
(\byear{1991})
\end{barticle}
%
\OrigBibText
\begin{barticle}
\bauthor{\bsnm{Tolley}, \binits{H.D.}},
\bauthor{\bsnm{Manton}, \binits{K.G.}}:
\batitle{Intervention effects among a collection of risks}.
\bjtitle{Transactions of the Society of Actuaries}
\bvolume{43},
\bfpage{443}--\blpage{467}
(\byear{1991})
\end{barticle}
\endOrigBibText
\bptok{structpyb}%
\endbibitem

\end{thebibliography}

\end{document}